\documentclass[]{gPAA2e}

\usepackage[bookmarks=false, hidelinks]{hyperref}

\usepackage{algorithmic}
\usepackage{algorithm}

\usepackage{multirow}

\usepackage{subcaption}

\begin{document}

\markboth{Adnan Ashraf and Ivan Porres}{Multi-objective dynamic virtual machine consolidation in the cloud using ant colony system}


\title{Multi-objective dynamic virtual machine consolidation in the cloud using ant colony system}

\author{Adnan Ashraf$^{\rm }$$^{\ast}$\thanks{$^\ast$Corresponding author. Email: adnan.ashraf@abo.fi
\vspace{6pt}} and Ivan Porres$^{\rm }$
\\\vspace{6pt}  $^{\rm }${\em{Faculty of Natural Sciences and Technology, \AA bo Akademi University, Finland}}
}

\maketitle

\begin{abstract}
In this paper, we present a novel multi-objective ant colony system algorithm for virtual machine (VM) consolidation in cloud data centers. The proposed algorithm builds VM migration plans, which are then used to minimize over-provisioning of physical machines (PMs) by consolidating VMs on under-utilized PMs. It optimizes two objectives that are ordered by their importance. The first and foremost objective in the proposed algorithm is to maximize the number of released PMs. Moreover, since VM migration is a resource-intensive operation, it also tries to minimize the number of VM migrations. The proposed algorithm is empirically evaluated in a series of experiments. The experimental results show that the proposed algorithm provides an efficient solution for VM consolidation in cloud data centers. Moreover, it outperforms two existing ant colony optimization based VM consolidation algorithms in terms of number of released PMs and number of VM migrations.

\bigskip

\begin{keywords}
virtual machines; consolidation; metaheuristic; ant colony system; cloud computing
\end{keywords}
\bigskip

\end{abstract}

\section{Introduction}

Cloud data centers comprise thousands of physical machines (PMs) and networking devices. These resources and their cooling infrastructure incur huge energy footprints. High energy consumption not only translates into a high operating cost, but also leads to huge carbon emissions~\cite{Kaur:2015:EET, Mastelic:2014}. Therefore, energy footprint of data centers is a major concern for cloud providers. The high energy consumption of data centers can partly be attributed to the large-scale installations of computing and cooling infrastructures, but more importantly it is due to the inefficient use of the computing resources~\cite{BeloglazovBuyya2012}.

Hardware virtualization technologies allow to share a PM among multiple, performance-isolated virtual machines (VMs) to improve resource utilization. Further improvement in resource utilization and reduction in energy consumption can be achieved by consolidating VMs on PMs and switching idle PMs off or to a low-power mode. VM consolidation has emerged as one of the most effective and promising techniques to reduce energy footprint of cloud data centers~\cite{BeloglazovBuyya2012, Farahnakian2015TSC}. A VM consolidation approach uses live VM migration to consolidate VMs on a reduced set of PMs. Thereby, allowing some of the underutilized PMs to be turned-off or switched to a low-power mode to conserve energy.

There is currently an increasing amount of interest on developing and evaluating efficient VM consolidation approaches for cloud data centers. Over the past few years, researchers have used a multitude of ways to develop novel VM consolidation approaches~\cite{BeloglazovBuyya2012, Beloglazov2012, Corradi:2014, Ferreto:2011, He:2012, Hwang:2013, Liao:2012, Murtazaev2011, Marzolla2011, Wang:2011, Wood:2009, Vogels2008}. Some of these approaches have also been reported in recent literature reviews~\cite{Ahmad:2015, Pires:2015}.

The main output of a VM consolidation algorithm is a VM migration plan, which is implemented by first migrating VMs from one PM to another and then shutting down or allocating new work to the idle PMs. The quality of a VM consolidation algorithm can be evaluated according to different criteria, including the number of released PMs (to be maximized), the number of VM migrations from one PM to another (to be minimized), and the algorithm execution time (to be minimized). Moreover, since a migration plan with a higher number of released PMs is always preferred to a migration plan with a lower number of VM migrations, maximizing the number of released PMs takes precedence over minimizing the number of VM migrations.

The VM consolidation problem is an NP-hard combinatorial optimization problem~\cite{Farahnakian2014CLOUD}. Therefore, it requires advanced strategies in order to be viable in practice. One way to address this problem is to use the exact optimization techniques, such as mixed-integer linear programming, which find optimal solutions with exponential runtime complexity. However, such techniques are mostly impractical for realistic, large-sized problem instances. Moreover, the currently available commercial, exact optimization tools, such as IBM ILOG CPLEX\footnote{\url{www.ibm.com/software/commerce/optimization/cplex-optimizer}}, do not provide support for multi-objective optimization problems. A widely-used alternative approach to solve difficult combinatorial optimization problems involves the use of metaheuristics~\cite{Blum2011}. Metaheuristics are high-level procedures that efficiently explore the search-space of available solutions with the aim to find near-optimal solutions with a polynomial time complexity~\cite{Harman2013}.

Some of the recent works~\cite{Farahnakian2014CLOUD, Feller2012CloudCom, Ferdaus2014, Farahnakian2015TSC} on VM consolidation use a highly adaptive metaheuristic called ant colony optimization (ACO)~\cite{Dorigo1999}. The existing ACO-based VM consolidation approaches~\cite{Farahnakian2014CLOUD, Feller2012CloudCom, Ferdaus2014, Farahnakian2015TSC} tend to use single-objective, single-colony algorithms with an aggregate objective function (AOF) that tries to combine multiple objectives. The benefit of the AOF approach is that it reduces complexity and may improve the runtime of the algorithm by limiting the search to a subspace of the feasible solutions. However, the main drawback is that a correct combination of the objectives requires certain weights to be assigned to each objective, which often requires an in-depth knowledge of the problem domain~\cite{Pires:2015}. Therefore, the assignment of the weights is essentially subjective~\cite{Hu:2013:Pareto}. Moreover, an AOF may not combine the optimization objectives in an appropriate manner. For instance, the AOFs in the existing ACO-based VM consolidation approaches~\cite{Farahnakian2014CLOUD, Feller2012CloudCom, Ferdaus2014, Farahnakian2015TSC} do not allow to order the objectives by their importance.

In this paper, we present a novel multi-objective ACO-based VM consolidation algorithm for cloud data centers that, according to our evaluation, outperforms the existing ACO-based VM consolidation algorithms~\cite{Farahnakian2014CLOUD, Feller2012CloudCom, Ferdaus2014, Farahnakian2015TSC}. It uses ant colony system (ACS)~\cite{Dorigo1997}, which is currently one of the best performing ACO algorithms. The proposed multi-objective ACS algorithm for VM consolidation is called MOACS and it optimizes two objectives. The first and foremost objective in MOACS is to maximize the number of released PMs. Moreover, since VM migration is a resource-intensive operation, MOACS also tries to minimize the number of VM migrations. We adapt and use a multi-objective, multi-colony ACS algorithm by Gambardella et al.~\cite{Gambardella1999MACS}, which orders the objectives by their importance. The proposed algorithm is not dependent on a particular deployment architecture or system topology. Therefore, it can be implemented in centralized as well as decentralized deployment architectures and system topologies, as the one proposed in~\cite{Feller2012CloudCom}.

\subsection{Contributions} 
\label{sec:overview}
The main task of the proposed MOACS algorithm is to find a VM migration plan that maximizes the number of released PMs while minimizing the number of VM migrations. In the rest of this section, we summarize the main contributions of this paper.

\subsubsection*{Improved Multi-Objective, Multi-Colony Optimization}
MOACS advances the state of the art on ACO-based VM consolidation by implementing a multi-objective, multi-colony ACS algorithm. It extends our previous single-objective, single-colony ACO algorithm for VM consolidation~\cite{Farahnakian2014CLOUD, Farahnakian2015TSC} and similar works by other researchers~\cite{Feller2012CloudCom, Ferdaus2014} that implement single-objective, single-colony ACO algorithms. The proposed multi-objective, multi-colony approach eliminates the need for an AOF and allows to combine the optimization objectives in an appropriate manner.

\subsubsection*{Improved Reduction of Search Space}
Since VM consolidation is an NP-hard problem, it requires fast and scalable algorithms. In order to improve the runtime performance of the proposed algorithm, we present three simple constraints in Section~\ref{sec:moacs_wac}. These constraints determine which PMs and VM migrations can be excluded from the consolidation process without compromising on the quality of the solutions. We refine two constraints from our previous work~\cite{Farahnakian2014CLOUD, Farahnakian2015TSC} and complement them with a new constraint concerning neighborhoods of PMs.

\subsubsection*{Improved Experimental Results}
We have implemented the proposed MOACS algorithm in Java and have compared it with two existing ACO-based VM consolidation algorithms. The first one is the single-objective, single-colony max-min ant system VM consolidation algorithm by Feller et al.~\cite{Feller2012CloudCom}, that we name Feller-ACO for evaluation.  We selected the Feller-ACO algorithm for comparison due to its excellent overall performance in many aspects as shown by its authors~\cite{Feller2012CloudCom}. The second one is our previously published single-objective, single-colony ACS VM consolidation algorithm~\cite{Farahnakian2015TSC}, that we refer to as ACS for evaluation. The ACS algorithm was selected as baseline for our work since it outperformed many other existing algorithms at the time of publication, as shown in ~\cite{Farahnakian2015TSC}. Our results show that the proposed MOACS algorithm outperforms Feller-ACO in all attributes: number of released PMs, number of VM migrations, packing efficiency, and algorithm execution time. Similarly, it outperforms ACS~\cite{Farahnakian2015TSC} in all attributes except in execution time.

We proceed as follows.  Section~\ref{related_work} provides background and discusses important related works. The proposed MOACS algorithm is described in detail in Section~\ref{sec:moacs_wac} while its experimental evaluation is presented in Section~\ref{experimental_design}.  Finally, we present our conclusions in Section~\ref{conclusion}.

\section{Background and Related Work}
\label{related_work}

Ant colony optimization (ACO) metaheuristic is inspired from the foraging behavior of real ant colonies~\cite{Dorigo1999}. While transporting food from the food source to their nest, ants deposit and follow trails of a chemical substance on their paths called pheromone. It allows them to indirectly communicate with one another to find better paths between their nest and the food source. Empirical results from previous research on ACO has shown that the simple pheromone trail following behavior of ants can give rise to the emergence of the shortest paths.

Each ant finds a complete path or solution, but high quality solutions emerge from the indirect communication and global cooperation among multiple concurrent ants~\cite{Farahnakian2015TSC}. Ants must also avoid stagnation, which is a premature convergence to a suboptimal solution. It is achieved by using pheromone evaporation and stochastic state transitions. There are a number of ACO algorithms, such as ant system (AS), ant colony system (ACS), and max-min ant system (MMAS)~\cite{Dorigo1999}. ACS~\cite{Dorigo1997} is currently one of the best performing ant algorithms. Therefore, in this paper, we apply ACS to the VM consolidation problem.

The existing ACO-based resource allocation, VM placement, and VM consolidation approaches include~\cite{Farahnakian2014CLOUD, Feller2012CloudCom, Ferdaus2014, Farahnakian2015TSC, ashraf2014pdp, Yin2006, Chaharsooghi:2008, Gao:2013}. Yin and Wang~\cite{Yin2006} applied ACO to the nonlinear resource allocation problem, which seeks to find an optimal allocation of a limited amount of resources to a number of tasks. Chaharsooghi and Kermani~\cite{Chaharsooghi:2008} proposed a modified version of ACO for the multi-objective resource allocation problem. Feller et al.~\cite{Feller2012CloudCom} applied the single-objective, single-colony MMAS algorithm to the VM consolidation problem in the context of cloud computing. Ferdaus et al.~\cite{Ferdaus2014} integrated ACS with a vector algebra-based server resource utilization capturing technique~\cite{Mishra2011}. In our previous work~\cite{ashraf2014pdp}, we applied the original single-objective, single-colony ACS algorithm~\cite{Dorigo1997} to the web application consolidation problem. Similarly, in our previous work~\cite{Farahnakian2014CLOUD, Farahnakian2015TSC}, we used the original ACS algorithm~\cite{Dorigo1997} for energy-aware VM consolidation in cloud data centers.

Gao et al.~\cite{Gao:2013} used a multi-objective ACS algorithm with two equally important objectives: minimize energy consumption and minimize resource wastage. In their approach, both energy consumption and resource wastage are derived from the number of PMs used for the placement of VMs. Their approach only provides an initial placement of VMs on PMs. It does not migrate VMs from one PM to another. Therefore, it can not be used to consolidate VMs or to minimize the number of VM migrations. The output of Gao et al.'s algorithm is a Pareto set of solutions, from which a solution is randomly selected. A drawback of this approach is that the objectives can not be ordered by their importance. Moreover, the randomly selected solution may not be the most desired solution. To the best of our knowledge, none of the existing ACO-based VM consolidation approaches use a multi-objective ACS algorithm that orders the objectives by their importance.

The main difference between the existing ACO-based VM consolidation approaches~\cite{Farahnakian2014CLOUD, Feller2012CloudCom, Ferdaus2014, Farahnakian2015TSC} and our proposed MOACS algorithm is that the existing approaches tend to use a single-objective, single-colony ACO algorithm with an AOF that tries to combine multiple objectives, whereas MOACS uses a multi-objective ACS algorithm with two independent ant colonies. The first colony maximizes the number of released PMs, while the second colony minimizes the number of VM migrations. We adapt and use a multi-objective, multi-colony ACS algorithm by Gambardella et al.~\cite{Gambardella1999MACS}, which was originally proposed for the vehicle routing problem with time windows.

\section{Multi-Objective ACS Algorithm for VM Consolidation} 
\label{sec:moacs_wac}

In this section, we present our proposed multi-objective ACS-based VM consolidation algorithm (MOACS). As its first objective, MOACS maximizes the number of released PMs $|P_{R}|$. Moreover, its second objective is to minimize the number of VM migrations $nM$. Since a global best migration plan $\Psi^{+}$ with a higher number of released PMs $|P_{R}|$ is always preferred to a $\Psi^{+}$ with a lower $|P_{R}|$ even if the number of VM migrations $nM$ is higher in the former $\Psi^{+}$, maximizing $|P_{R}|$ takes precedence over minimizing $nM$. For the sake of clarity, important concepts and notations used in the following sections are tabulated in Table~\ref{tab:notation}.

\begin{table}[!t]
\begin{center}
\caption{Summary of concepts and their notations}
\label{tab:notation}
\begin{tabular}{ll} 
\toprule

$M$ & set of migration plans \\
$P$ & set of PMs \\
$P_{R}$ & set of PMs that are released when a migration plan $\Psi$ is enforced \\
$T$ & set of tuples \\
$T_k$ & set of tuples not yet traversed by ant $k$ \\


$V$ & set of VMs \\
$V_p$ & set of VMs running on a PM $p$ \\

\hline
$C_{p_{de}}$ & total capacity vector of the destination PM $p_{de}$ \\
$N$ & a neighborhood of PMs \\
$p_{de}$ & destination PM in a tuple\\
$p_{so}$ & source PM in a tuple \\
$q$ & a uniformly distributed random variable \\
$S$ & a random variable selected according to~\eqref{state_transition_rule_2}\\ 
$Scr_k$ & thus far best score of ant $k$ \\

$U_{p_{de}}$ & used capacity vector of the destination PM $p_{de}$ \\
$U_{p_{so}}$ & used capacity vector of the source PM $p_{so}$ \\
$U_v$ & used capacity vector of the VM $v$ \\

\hline
$\eta$ & heuristic value \\
$\tau$ & amount of pheromone \\
$\tau_0$ & initial pheromone level \\

$\Psi$ & a migration plan \\
$\Psi^{+}$ & the global best migration plan \\
$\Psi^{+}_{nM}$ & thus far best migration plan from ACS$_{nM}$ \\
$\Psi^{+}_{PR}$ & thus far best migration plan from ACS$_{|PR|}$ \\
$\Psi_k$ & ant-specific migration plan of ant $k$ \\
$\Psi_k^m$ & ant-specific temporary migration plan of ant $k$ \\

$\Delta_{\tau_s}^{nM}$ & additional pheromone amount given to the tuples in $\Psi^{+}_{nM}$ \\
$\Delta_{\tau_s}^{PR}$ & additional pheromone amount given to the tuples in $\Psi^{+}_{PR}$ \\

\hline

$\alpha$  & pheromone decay parameter in the global updating rule \\
$\beta$ & parameter to determine the relative importance of $\eta$ \\
$\rho$  & pheromone decay parameter in the local updating rule \\
$q_0$  & parameter to determine relative importance of exploitation \\ 

\hline
$nA$ & number of ants that concurrently build their migration plans \\
$nI$ & number of ant generations \\
$nM$ & number of VM migrations \\




\hline
$f(\Psi)$ & objective function in~\eqref{objective_function} concerning number of released PMs \\
$g(\Psi)$ & objective function in~\eqref{objective_function_nm} concerning number of VM migrations \\

\botrule 
\end{tabular}
\end{center}
\end{table}

\begin{figure}[!ht]
	\centering	
    \includegraphics[width=0.90\textwidth]{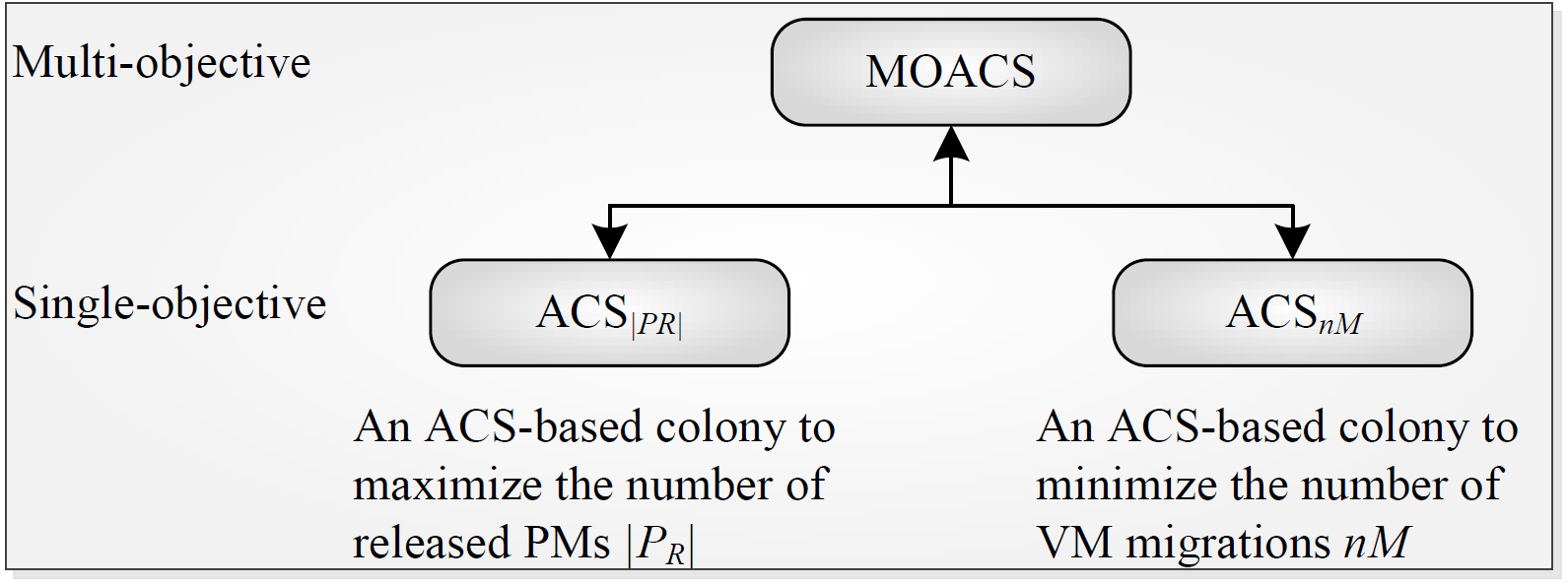}
	\caption{MOACS architecture}
	\label{fig:MOACSWACarchitecture}
\end{figure}

Figure~\ref{fig:MOACSWACarchitecture} illustrates MOACS architecture. MOACS coordinates the operations of two ACS-based ant colonies, which simultaneously optimize their respective objectives. The first objective concerning the number of released PMs $|P_{R}|$ is optimized by the first colony called ACS$_{|PR|}$. Similarly, the second colony called ACS$_{nM}$ optimizes the second objective concerning the number of VM migrations $nM$. Both colonies work independently and use independent pheromone and heuristic matrices. However, they collaborate on the global best migration plan $\Psi^{+}$, which is maintained by MOACS.

In the VM consolidation problem, each PM $p \in P$ hosts multiple VMs $v \in V$. Each PM that hosts at least one VM is a potential \emph{source PM}. Both the source PM and the VM are characterized by their resource utilizations, such as CPU utilization and memory utilization. MOACS uses a notion of neighborhoods of PMs. Neighborhoods are mutually exclusive subsets of $P$. A neighborhood is an abstract entity that represents a set of closely-located PMs in a cloud data center. For example, the PMs in a data center rack may constitute a neighborhood of PMs. A VM can be migrated to any other PM located in any neighborhood $N$. Therefore, every other PM within as well as outside the neighborhood of the source PM is a potential \emph{destination PM}, which is also characterized by its resource utilizations. Thus, MOACS makes a set of tuples $T$, where each tuple $t \in T$ consists of three elements: source PM $p_{so}$, VM $v$, and destination PM $p_{de}$
\begin{equation}
\label{tuple}
  t := (p_{so}, v, p_{de})
\end{equation}

The computation time of the proposed VM consolidation algorithm is primarily based on the number of tuples $|T|$. Therefore, in order to reduce the computation time, MOACS applies three constraints, which result in a reduced set of tuples by removing some least important and unwanted tuples. The first constraint ensures that only under-utilized PMs are used as the source PMs. Similarly, the second constraint allows only under-utilized PMs to be considered as the destination PMs. In other words, migrations from and to well-utilized PMs are excluded. The rationale is that a well-utilized PM should not become part of the consolidation process because migrating to a well-utilized PM may result in its overloading. Similarly, migrating from a well-utilized PM is less likely to result in the termination of the source PM and thus it would not reduce the total number of required PMs. The third constraint further restricts the size of the set of tuples $|T|$ by preventing inter-neighborhood migrations. Therefore, as a general rule, a VM can only be migrated to another PM within the neighborhood of its source PM. The only exception to this rule is when a neighborhood has only one PM in it. In this case, the VMs from the lone PM can be migrated to any other PM in any neighborhood. By applying these three simple constraints in a series of preliminary experiments, we observed that the computation time of the algorithm was significantly reduced without compromising on the quality of the solutions.

The space complexity of the proposed algorithm is $\mathcal{O}(|T|)$, where $|T|$ is the number of tuples. Moreover, the worst-case space complexity corresponds to a VM consolidation scenario that does not involve any well-utilized PMs. In such a scenario, each PM is considered as a source as well as a destination PM. The maximum number of tuples in the worst-case is computed as
\begin{equation}
\label{number_of_tuples}
  maximum \; |T| := |P| \cdot |V| \cdot (|N|-1)
\end{equation}
where $|P|$ is the number of PMs, $|V|$ is the number of VMs, and $|N|$ is the neighborhood size. Since real VM consolidation scenarios usually involve one or more well-utilized PMs and the proposed algorithm excludes migrations from and to well-utilized PMs, the actual number of tuples $|T|$ in a real scenario is often smaller than that of the worst-case scenario.

The pseudocode of the proposed MOACS algorithm is presented in Algorithm~\ref{algo:moacs_wac}. Initially, the global best migration plan $\Psi^{+}$ is not known. Therefore, $\Psi^{+}$ is empty (line~2). The main loop in line~1 iterates until a stopping criterion is met. For instance, when all remaining PMs are well-utilized or when no further improvements are achieved in a given number of consecutive iterations~\cite{Gambardella1999MACS}. In each iteration of the main loop, the two ACS-based colonies ACS$_{|PR|}$ and ACS$_{nM}$ try to find the global best migration plan $\Psi^{+}$ according to their respective objectives. ACS$_{|PR|}$ tries to find a migration plan with a higher number of released PMs $|P_{R}|$ (line~3). Similarly, ACS$_{nM}$ tries to find a migration plan with fewer VM migrations (line~4). The global best migration plan $\Psi^{+}$ is updated every time an improved migration plan is found. If ACS$_{|PR|}$ finds a migration plan with a higher number of released PMs (line~5), $\Psi^{+}$ is updated according to the thus far best migration plan from ACS$_{|PR|}$ denoted as $\Psi^{+}_{PR}$ (line~6). Likewise, when ACS$_{nM}$ finds a migration plan with fewer VM migrations, but with at least as many released PMs $P_{R}$ as in $\Psi^{+}$ (lines~8--9), $\Psi^{+}$ is updated according to the thus far best migration plan from ACS$_{nM}$ denoted as $\Psi^{+}_{nM}$ (line~10). Finally, at the end of each iteration of the main loop, VMs are consolidated according to the global best migration plan $\Psi^{+}$ and the released PMs are terminated (line~13).

\begin{algorithm}[!b]
\caption{Multi-objective ACS algorithm for VM consolidation (MOACS)}
\label{algo:moacs_wac}
\begin{algorithmic}[1]

\WHILE {until a stopping criterion is met}

\STATE $\Psi^{+} := \emptyset$ 

\STATE $\Psi^{+}_{PR} := ACS_{|PR|}$ 
\STATE $\Psi^{+}_{nM} := ACS_{nM}$ 
\IF {$\Psi^{+} = \emptyset \vee f(\Psi^{+}_{PR}) > f(\Psi^{+})$} 
\STATE $\Psi^{+} := \Psi^{+}_{PR}$
\ENDIF
\IF {$f(\Psi^{+}_{nM}) \geq f(\Psi^{+})$} 
\IF {$g(\Psi^{+}_{nM}) > g(\Psi^{+})$} 
\STATE $\Psi^{+} := \Psi^{+}_{nM}$
\ENDIF
\ENDIF

\STATE consolidate VMs according to $\Psi^{+}$ and terminate released PMs

\ENDWHILE
\end{algorithmic}
\end{algorithm}

\subsection{ACS-based Colony to Maximize the Number of Released PMs}

The ACS$_{|PR|}$ colony optimizes the first objective concerning the number of released PMs $|P_{R}|$. Therefore, the objective function for the ACS$_{|PR|}$ algorithm is
\begin{equation}
\label{objective_function}
  maximize \; f(\Psi):= |P_{R}|
\end{equation}
where $\Psi$ is the migration plan and $P_{R}$ is the set of PMs that will be released when $\Psi$ is enforced. Since the primary objective of VM consolidation is to minimize the number of active PMs, the objective function is defined in terms of number of released PMs $|P_{R}|$. Moreover, when a migration plan is enforced, we apply a constraint which reduces the number of VM migrations $nM$ by restricting migrations to only those PMs that are not included in the set of released PMs $P_{R}$, that is
\begin{equation}
\label{destinationVM_constraint}
    \forall \; p_{de} \; \in \; P \; | \; p_{de} \; \notin \;P_{R}
\end{equation}

In our approach, a PM can only be considered as released when all VMs migrate from it. Therefore, the set of released PMs $P_{R}$ is defined as
\begin{equation}
\label{released_PMs}
  P_{R} := \{\forall p \; \in \; P | V_p = \emptyset \}
\end{equation}
where $V_p$ is the set of VMs running on a PM $p$. Thus, a PM can only be included in the set of released PMs $P_{R}$ when it no longer hosts any VMs.

Since there is no notion of path in the VM consolidation problem, ants deposit pheromone on the tuples defined in~\eqref{tuple}. Each of the $nA$ ants uses a stochastic state transition rule to choose the next tuple to traverse. The state transition rule in ACS$_{|PR|}$ is called pseudo-random-proportional-rule~\cite{Dorigo1997}. According to this rule, an ant $k$ chooses a tuple $s$ to traverse next by applying
\begin{equation}
\label{state_transition_rule_1}
  s:=
  \begin{cases}
  \text{arg} \; \text{max}_{u \; \in \; T_k} \{[\tau_u] \cdot [\eta_u]^{\beta}\}, & \text{if } q \leq q_0
  \\
  S, & \text{otherwise}
  \end{cases}
\end{equation}
where $\tau$ denotes the amount of pheromone and $\eta$ represents the heuristic value associated with a particular tuple. $\beta$ is a parameter to determine the relative importance of the heuristic value with respect to the pheromone value. The expression \emph{arg max} returns the tuple for which $[\tau] \cdot [\eta]^{\beta}$ attains its maximum value. $T_k \subset T$ is the set of tuples that remain to be traversed by ant $k$. $q \in [0,1]$ is a uniformly distributed random variable and $q_0 \in [0,1]$ is a parameter. $S$ is a random variable selected according to the probability distribution given in~\eqref{state_transition_rule_2}, where the probability $prob_s$ of an ant $k$ to choose tuple $s$ to traverse next is defined as
\begin{equation}
\label{state_transition_rule_2}
  prob_s:=
  \begin{cases}
  \frac {[\tau_s] \cdot [\eta_s]^{\beta}} {\sum\limits_{u \; \in \; T_k} [\tau_u] \cdot [\eta_u]^{\beta}}, & \text{if } s \; \in \; T_k
  \\
  0, & \text{otherwise}
  \end{cases}
\end{equation}
The heuristic value $\eta_s$ of a tuple $s$ is defined as
\begin{equation}
\label{heuristic_value}
  \eta_s :=
  \begin{cases}
  \frac {U_{p_{de}} + U_v} {C_{p_{de}}}, & \text{if } U_{p_{de}} + U_v \leq C_{p_{de}}
  \\
  0, & \text{otherwise}
  \end{cases}
\end{equation}
where $C_{p_{de}}$ is the total capacity vector of the destination PM $p_{de}$, $U_{p_{de}}$ is the used capacity vector of $p_{de}$, and likewise $U_v$ is the used capacity vector of the VM $v$ in tuple $s$. The heuristic value $\eta$ is based on the ratio of $(U_{p_{de}} + U_v)$ to $C_{p_{de}}$. Therefore, destination PMs with the minimum unused capacity receive the highest amount of heuristic value. Thus, the heuristic value favors VM migrations that would result in a reduced under-utilization of PMs. Moreover, the constraint $U_{p_{de}} + U_v \leq C_{p_{de}}$ prevents VM migrations that would result in the overloading of the destination PM $p_{de}$. In the proposed algorithm, we assumed two resource dimensions, which represent CPU utilization and memory utilization. However, if necessary, it is possible to add more dimensions in the total and used capacity vectors.

The stochastic state transition rule in~\eqref{state_transition_rule_1} and~\eqref{state_transition_rule_2} prefers tuples with a higher pheromone concentration and which result in a higher number of released PMs. The first case in~\eqref{state_transition_rule_1} where $q \leq q_0$ is called exploitation~\cite{Dorigo1997}. It chooses the best tuple that attains the maximum value of $[\tau] \cdot [\eta]^{\beta}$. The second case, called biased exploration, selects a tuple according to~\eqref{state_transition_rule_2}. The exploitation helps the ants to quickly converge to a high quality solution, while at the same time, the biased exploration helps them to avoid stagnation by allowing a wider exploration of the search space. In addition to the stochastic state transition rule, ACS$_{|PR|}$ also uses a global and a local pheromone trail evaporation rule. The global pheromone trail evaporation rule is applied towards the end of an iteration after all ants complete their migration plans. It is defined as
\begin{equation}\label{global_updating_rule_vr}
  \tau_s := (1-\alpha) \cdot \tau_s + \alpha \cdot \Delta_{\tau_s}^{PR}
\end{equation}
where $\alpha \in (0,1]$ is the pheromone decay parameter and $\Delta_{\tau_s}^{PR}$ is the additional pheromone amount that is given only to those tuples that belong to the thus far best migration plan from ACS$_{|PR|}$ denoted as $\Psi^{+}_{PR}$ in order to reward them. It is defined as
\begin{equation}
\label{delta_tau_vr}
  \Delta_{\tau_s}^{PR} :=
  \begin{cases}
  |P_{R}|, & \text{if } s \; \in \; \Psi^{+}_{PR}
  \\
  0, & \text{otherwise}
  \end{cases}
\end{equation}

The local pheromone trail update rule is applied on a tuple when an ant traverses the tuple while making its migration plan. It is defined as
\begin{equation}\label{local_updating_rule}
  \tau_s := (1-\rho) \cdot \tau_s + \rho \cdot \tau_0
\end{equation}
where $\rho \in (0,1]$ is similar to $\alpha$ and $\tau_0$ is the initial pheromone level, which is computed as the multiplicative inverse of the product of the number of PMs $|P|$ and the approximate optimal $|\Psi|$
\begin{equation}\label{tau_0}
  \tau_0 := (|\Psi| \cdot |P|)^{-1}
\end{equation}
Here, any very rough approximation of the optimal $|\Psi|$ suffices~\cite{Dorigo1997}. The pseudo-random-proportional-rule in ACS$_{|PR|}$ and the global pheromone trail update rule are intended to make the search more directed. The pseudo-random-proportional-rule prefers tuples with a higher pheromone level and a higher heuristic value. Therefore, the ants try to search other high quality solutions in the close proximity of the thus far global best solution. On the other hand, the local pheromone trail update rule complements exploration of other high quality solutions that may exist far form the thus far global best solution. This is because whenever an ant traverses a tuple and applies the local pheromone trail update rule, the tuple looses some of its pheromone and thus becomes less attractive for other ants. Therefore, it helps in avoiding stagnation where all ants end up finding the same solution or where they prematurely converge to a suboptimal solution.

The pseudocode of the ACS$_{|PR|}$ algorithm is given as Algorithm~\ref{algo:moacs_vr}. The algorithm makes a set of tuples $T$ using~\eqref{tuple} and sets the pheromone value of each tuple to the initial pheromone level $\tau_0$ by using~\eqref{tau_0} (line~2). Then, it iterates over $nI$ iterations (line~3), where each iteration $i \in nI$ creates a new generation of $nA$ ants that concurrently build their migration plans (lines~4--20). Each ant $k \in nA$ iterates over $|T|$ tuples (lines~6--18). It computes the probability of choosing the next tuple to traverse by using~\eqref{state_transition_rule_2} (line~7). Afterwards, based on the computed probabilities and the stochastic state transition rule in~\eqref{state_transition_rule_1} and~\eqref{state_transition_rule_2}, each ant chooses a tuple $t$ to traverse (line~8) and adds $t$ to its temporary migration plan $\Psi_k^m$ (line~9). The local pheromone trail update rule in~\eqref{local_updating_rule} and~\eqref{tau_0} is applied on $t$ (line~10). If the migration in $t$ does not overload the destination PM $p_{de}$, the used capacity vectors at the source PM $U_{p_{so}}$ and the destination PM $U_{p_{de}}$ in $t$ are updated to reflect the impact of the migration (line~12). Then, the objective function in~\eqref{objective_function} is applied on $\Psi_k^m$. If it yields a score higher than the ant's thus far best score $Scr_k$ (line~13), $t$ is added to the ant-specific migration plan $\Psi_k$ (line~15). Afterwards, when all ants complete their migration plans, all ant-specific migration plans are added to the set of migration plans $M$ (line~19), each migration plan $\Psi_k \in M$ is evaluated by applying the objective function in~\eqref{objective_function}, the thus far global best VM migration plan $\Psi^{+}$ is selected (line~21), and the global pheromone trail update rule in~\eqref{global_updating_rule_vr} and~\eqref{delta_tau_vr} is applied on all tuples (line~22). Finally, when all iterations $i \in nI$ complete, ACS$_{|PR|}$ returns the thus far best migration plan from ACS$_{|PR|}$ denoted as $\Psi^{+}_{PR}$ (line~24).

\begin{algorithm}[t]
\caption{ACS-based colony to maximize the number of released PMs (ACS$_{|PR|}$)}
\label{algo:moacs_vr}
\begin{algorithmic}[1]
\STATE $\Psi^{+}_{PR} := \emptyset$, $M := \emptyset$ 
\STATE $ \forall t \; \in \; T | \tau_t := \tau_0 $ 
\FOR{$i \; \in \; [1,nI] $} 
\FOR{$k \; \in \; [1,nA] $} 
\STATE $\Psi_k^m:=\emptyset, \Psi_k:=\emptyset, Scr_k:=0$
\WHILE{$|\Psi_k^m| < |T|$} 
\STATE compute $prob_s$ \; $\forall s \; \in \; T$ by using~\eqref{state_transition_rule_2}
\STATE choose a tuple $t \; \in \; T$ to traverse by using~\eqref{state_transition_rule_1}
\STATE $\Psi_k^m := \Psi_k^m \cup \{t\}$ 
\STATE apply local update rule in~\eqref{local_updating_rule} on $t$ 

\IF {the migration in $t$ does not overload destination PM $p_{de}$} 

\STATE update used capacity vectors $U_{p_{so}}$ and $U_{p_{de}}$ in $t$
\IF {$f(\Psi_k^m) > Scr_k$} %
\STATE $Scr_k := f(\Psi_k^m)$
\STATE $\Psi_k := \Psi_k \cup \{t\}$
\ENDIF

\ENDIF

\ENDWHILE
\STATE $M := M \cup \{\Psi_k\}$
\ENDFOR
\STATE $\Psi^{+}_{PR} := \text{arg} \; \text{max}_{\Psi_k \; \in \; M} \{ f(\Psi_k) \}$ 
\STATE apply global update rule in~\eqref{global_updating_rule_vr} on all $s \; \in \; T$ 
\ENDFOR
\STATE return $\Psi^{+}_{PR}$
\end{algorithmic}
\end{algorithm}

The time complexity of the ACS$_{|PR|}$ algorithm is $\mathcal{O}(nI \cdot |T|^2)$, where $nI$ is the number of ant generations and $|T|$ is the number of tuples. It can be derived from the pseudocode in Algorithm~\ref{algo:moacs_vr}. The main loop in line~3 iterates over $nI$. The second loop in line~4 does not add to the time complexity because the ants concurrently build their migration plans. The while loop in line~6 iterates over $|T|$. Finally, the probability calculation in line~7 requires an iteration over $|T|$.

\subsection{ACS-based Colony to Minimize the Number of VM Migrations}
The ACS$_{nM}$ algorithm tries to find a migration plan with fewer VM migrations, but with at least as many released PMs $P_{R}$ as in $\Psi^{+}$. Thus, the objective function for ACS$_{nM}$ is
\begin{equation}
\label{objective_function_nm}
  maximize \; g(\Psi):= (nM)^{-1} 
\end{equation}
where $\Psi$ is the migration plan and $nM$ is the number of VM migrations. Since VM migration is a resource-intensive operation, the objective function for ACS$_{nM}$ is defined as the multiplicative inverse of the number of VM migrations $nM$.

The ants in the ACS$_{nM}$ colony use the same pseudo-random-proportional-rule as in~\eqref{state_transition_rule_1} and~\eqref{state_transition_rule_2} to choose the next tuple to traverse. Moreover, as a general rule, the heuristic value $\eta_s$ in~\eqref{heuristic_value} favors tuples with a greater VM used capacity vector $U_v$. Therefore, the VM migrations that are more likely to result in a reduced number of VM migrations receive a higher amount of heuristic value $\eta_s$. Thus, the heuristic value $\eta_s$ in~\eqref{heuristic_value} supports the objective function of ACS$_{nM}$ in~\eqref{objective_function_nm}.

The ACS$_{nM}$ colony also uses the same local pheromone trail update rule as in~\eqref{local_updating_rule}. However, the global pheromone trail evaporation rule in ACS$_{nM}$ is defined as
\begin{equation}\label{global_updating_rule_nm}
  \tau_s := (1-\alpha) \cdot \tau_s + \alpha \cdot \Delta_{\tau_s}^{nM}
\end{equation}
where $\Delta_{\tau_s}^{nM}$ is the additional pheromone amount that is given only to those tuples that belong to the thus far best migration plan from ACS$_{nM}$ denoted as $\Psi^{+}_{nM}$ in order to reward them. It is defined as
\begin{equation}
\label{delta_tau_nm}
  \Delta_{\tau_s}^{nM} :=
  \begin{cases}
  (nM)^{-1}, & \text{if } s \; \in \; \Psi^{+}_{nM}
  \\
  0, & \text{otherwise}
  \end{cases}
\end{equation}

The pseudocode of the ACS$_{nM}$ algorithm is given as Algorithm~\ref{algo:moacs_nm}. Most of the steps in ACS$_{nM}$ are similar to those in the ACS$_{|PR|}$ colony in Algorithm~\ref{algo:moacs_vr} (lines~1--12). 
Similarly, in line~13, the algorithm uses the objective function concerning the number of released PMs $|P_{R}|$ defined in~\eqref{objective_function} instead of the objective function concerning the number of VM migrations $nM$ defined in~\eqref{objective_function_nm} because at this step it is important to find a migration plan with a higher number of released PMs $|P_{R}|$. 
However, when selecting the thus far best migration plan from ACS$_{nM}$ denoted as $\Psi^{+}_{nM}$ (line~21), all ant-specific migration plans $\Psi_k \in M$ are evaluated first by applying the objective function concerning the number of released PMs $|P_{R}|$ defined in~\eqref{objective_function} and then by the objective function concerning the number of VM migrations $nM$ defined in~\eqref{objective_function_nm}. Therefore, the migration plan with the highest number of released PMs $|P_{R}|$ and a lower number of VM migrations $nM$ is selected as the best migration plan $\Psi^{+}_{nM}$. Afterwards, the algorithm applies the global pheromone trail update rule of the ACS$_{nM}$ colony defined in~\eqref{global_updating_rule_nm} and~\eqref{delta_tau_nm} on all tuples (line~22). Finally, it returns $\Psi^{+}_{nM}$ (line~24).

The time complexity of the ACS$_{nM}$ algorithm is similar to that of the ACS$_{|PR|}$ algorithm. Since the ACS$_{|PR|}$ and ACS$_{nM}$ colonies work concurrently and independently, the overall time complexity of the proposed MOACS algorithm for finding the global best migration plan $\Psi^{+}$ is $\mathcal{O}(nI \cdot |T|^2)$.

\begin{algorithm}[t]
\caption{ACS-based colony to minimize the number of VM migrations (ACS$_{nM}$)}
\label{algo:moacs_nm}
\begin{algorithmic}[1]
\STATE $\Psi^{+}_{nM} := \emptyset$, $M := \emptyset$ 
\STATE $ \forall t \; \in \; T | \tau_t := \tau_0 $ 
\FOR{$i \; \in \; [1,nI] $} 
\FOR{$k \; \in \; [1,nA] $} 

\STATE $\Psi_k^m:=\emptyset, \Psi_k:=\emptyset, Scr_k:=0$ 


\WHILE{$|\Psi_k^m| < |T|$} 

\STATE compute $prob_s$ \; $\forall s \; \in \; T$ by using~\eqref{state_transition_rule_2}
\STATE choose a tuple $t \; \in \; T$ to traverse by using~\eqref{state_transition_rule_1}

\STATE $\Psi_k^m := \Psi_k^m \cup \{t\}$ 

\STATE apply local update rule in~\eqref{local_updating_rule} on $t$ 


\IF {the migration in $t$ does not overload destination PM $p_{de}$} 

\STATE update used capacity vectors $U_{p_{so}}$ and $U_{p_{de}}$ in $t$

\IF {$f(\Psi_k^m) > Scr_k$} 
\STATE $Scr_k := f(\Psi_k^m)$ 

\STATE $\Psi_k := \Psi_k \cup \{t\}$

\ENDIF

\ENDIF


\ENDWHILE
\STATE $M := M \cup \{\Psi_k\}$
\ENDFOR

\STATE $\Psi^{+}_{nM} := \text{arg} \; \text{max}_{\Psi_k \; \in \; M} \{ f(\Psi_k) \} \wedge \text{ arg} \; \text{max}_{\Psi_k \; \in \; M} \{ g(\Psi_k) \}$


\STATE apply global update rule in~\eqref{global_updating_rule_nm} on all $s \; \in \; T$ 
\ENDFOR
\STATE return $\Psi^{+}_{nM}$
\end{algorithmic}
\end{algorithm}

\section{Evaluation}
\label{experimental_design}

In this section, we describe the experimental evaluation of the proposed MOACS algorithm and its comparison with the single-objective, single-colony MMAS VM consolidation algorithm (Feller-ACO) by Feller et al.~\cite{Feller2012CloudCom} and our previously published single-objective, single-colony ACS VM consolidation algorithm (ACS)~\cite{Farahnakian2015TSC}.

We have implemented our proposed MOACS algorithm as a Java program called the MOACS Solver. It is available online under an open-source license\footnote{https://github.com/SELAB-AA/moacs-wac}. We have also developed Java solvers for the Feller-ACO~\cite{Feller2012CloudCom} and ACS~\cite{Farahnakian2015TSC} algorithms.

\subsection{Experimental Design}

The objective of the experiment was to compare the performance of the three implemented solvers: ACS, MOACS, and Feller-ACO. The input of these algorithms is a VM consolidation problem that can be characterized by the following parameters: number of PMs, number of VMs to consolidate, CPU utilization of each VM, memory requirements of each VM, and the current location of each VM. We used a factorial experiment design~\cite{Wohlin:2012}, in which the three solvers were tested in four different scenarios: (1) low CPU and small  memory requirements with respect to the capacity of the PMs, (2) high CPU and large memory requirements, (3) high CPU and small memory requirements, and (4) low CPU and large memory requirements. The experiment used randomly generated workloads, homogeneous VMs, and homogeneous PMs. The experimental parameters are summarized in Table~\ref{tab:design} and Table~\ref{tab:ACSParams}. For Scenario 1, the number of VMs to consolidate was 1000 and the number of PMs was 100 (ratio 10:1), while in the other 3 scenarios there were 1000 VMs and 200 PMs (ratio 5:1). The neighborhood size was set to 5 and the neighbors were chosen randomly. The ACO parameters used in the ACS, MOACS, and Feller-ACO solvers are tabulated in Table~\ref{tab:ACSParams}. These parameter values were obtained in a series of preliminary experiments. The dependent variables of the experiment were:
\begin{itemize}
\item Number of released PMs after consolidation, to be maximized.
\item Packing efficiency, defined as the ratio between the number of released PMs and the total number of PMs, to be maximized.
\item Number of VM migrations during consolidation, to be minimized.
\item Solver execution time, to be minimized.
\end{itemize}

\begin{table}[]
\centering
\caption{Experiment design}
\label{tab:design}
\begin{tabular}{llll}
\toprule
\multicolumn{2}{l}{\multirow{2}{*}{\begin{tabular}[c]{@{}l@{}}Algorithm\\ ACS, MOACS, Feller-ACO \end{tabular}}} & \multicolumn{2}{c}{CPU Utilization}                                                                                                                \\ \cline{3-4}
\multicolumn{2}{l}{}                                                                                          & \multicolumn{1}{c}{Low}                                                  & \multicolumn{1}{c}{High}                                               \\ \hline
\multicolumn{1}{c}{\multirow{6}{*}{Memory}}                              & Small                              & \begin{tabular}[c]{@{}l@{}}Scenario 1\\  $|V|$=1000\\  $|P|$=100\\ $|N|$=5\\ Number of runs=10\end{tabular} & \begin{tabular}[c]{@{}l@{}}Scenario 3\\ $|V|$=1000\\ $|P|$=200\\ $|N|$=5\\ Number of runs=10\end{tabular} \\ \cline{2-4}
\multicolumn{1}{c}{}                                                     & Large                              & \begin{tabular}[c]{@{}l@{}}Scenario 4\\ $|V|$=1000\\ $|P|$=200\\ $|N|$=5\\ Number of runs=10\end{tabular}   & \begin{tabular}[c]{@{}l@{}}Scenario 2\\ $|V|$=1000\\ $|P|$=200 \\ $|N|$=5\\ Number of runs=10\end{tabular} \\
\botrule
\end{tabular}

\vspace{0.4cm}

\caption{ACO parameters}
\label{tab:ACSParams}
\begin{tabular}{cccccc}
\toprule
$\alpha$ & $\beta$ & $\rho$ & $q_0$ & $nA$ & $nI$ \\
\hline
$0.1$ & $2.0$ & $0.1$ & $0.9$ & $10$ & $2$  \\
\botrule
\end{tabular}
\end{table}

\subsection{Execution}
The three solvers under evaluation used approximated algorithms. Therefore, we ran each solver ten times for each scenario, every time with a different random seed. Consequently, the experiment comprised a total of 40 test runs for each solver. The experiments were run on an Intel Core i7-4790 processor with 16 gigabytes of memory.

\subsection{Results}
\label{experimental_results}

\subsubsection{Number of Released PMs and Packing Efficiency}
\begin{figure}[t!]
    \centering
    \begin{subfigure}[t]{0.45\textwidth}
        \centering
        \includegraphics[trim=0 0 0 0, clip, width=\textwidth]{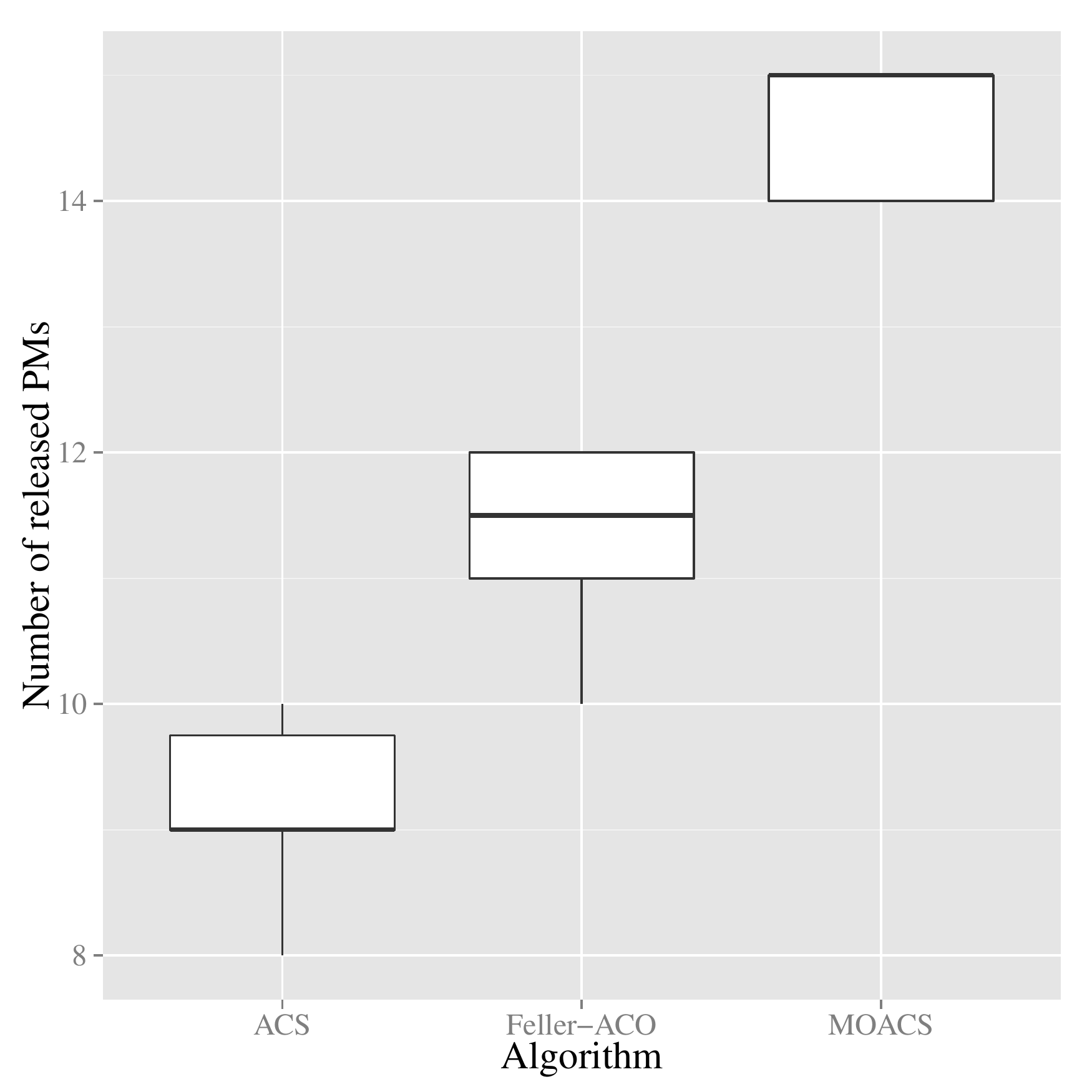}
        \caption{S1: Low CPU, small memory}
    \end{subfigure}
    \begin{subfigure}[t]{0.45\textwidth}
        \centering
        \includegraphics[trim=0 0 0 0, clip, width=\textwidth]{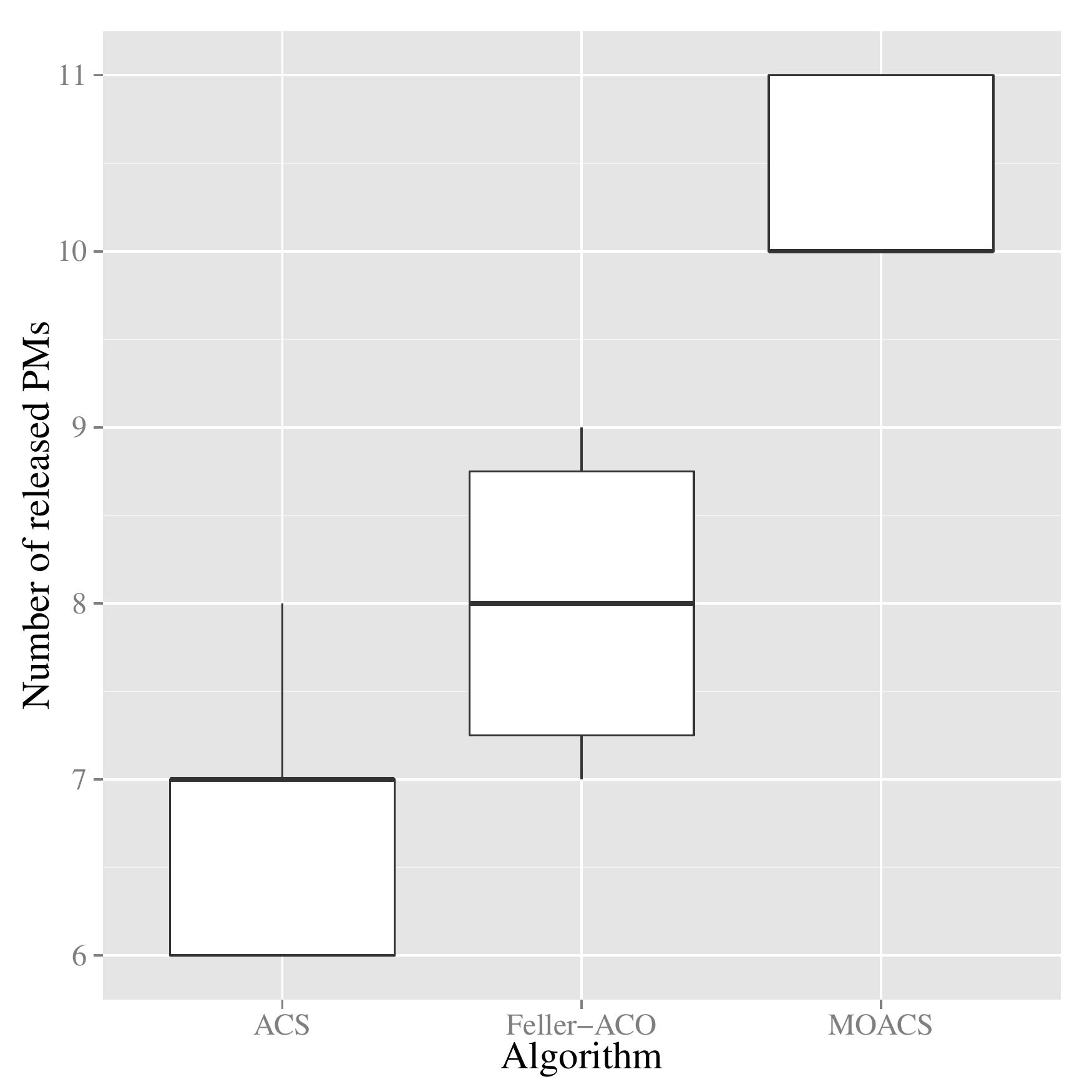}
        \caption{S3: High CPU, small memory}
    \end{subfigure}

    \begin{subfigure}[t]{0.45\textwidth}
        \centering
        \includegraphics[trim=0 0 0 0, clip, width=\textwidth]{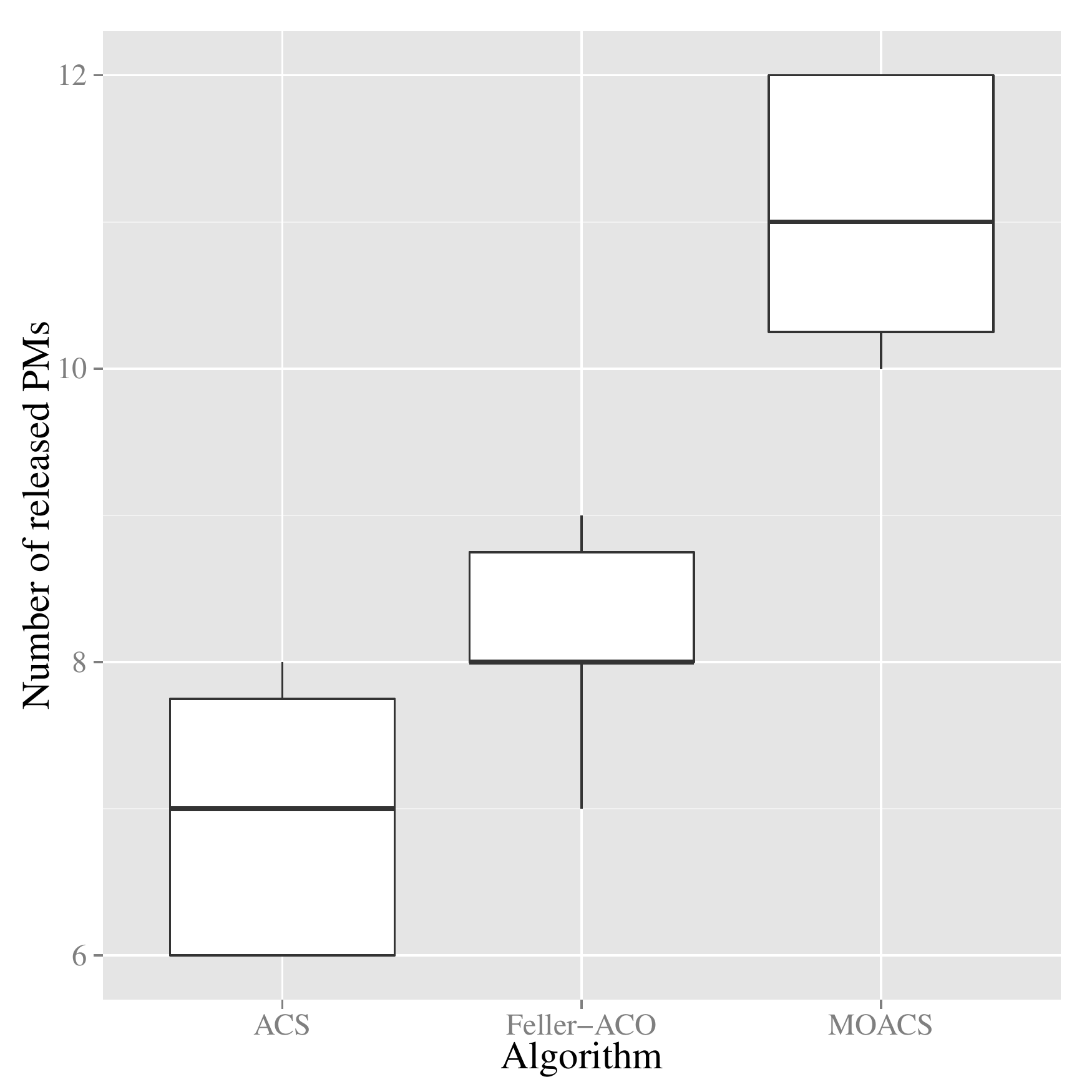}
        \caption{S4: Low CPU, large memory}
    \end{subfigure}
    \begin{subfigure}[t]{0.45\textwidth}
        \centering
        \includegraphics[trim=0 0 0 0, clip, width=\textwidth]{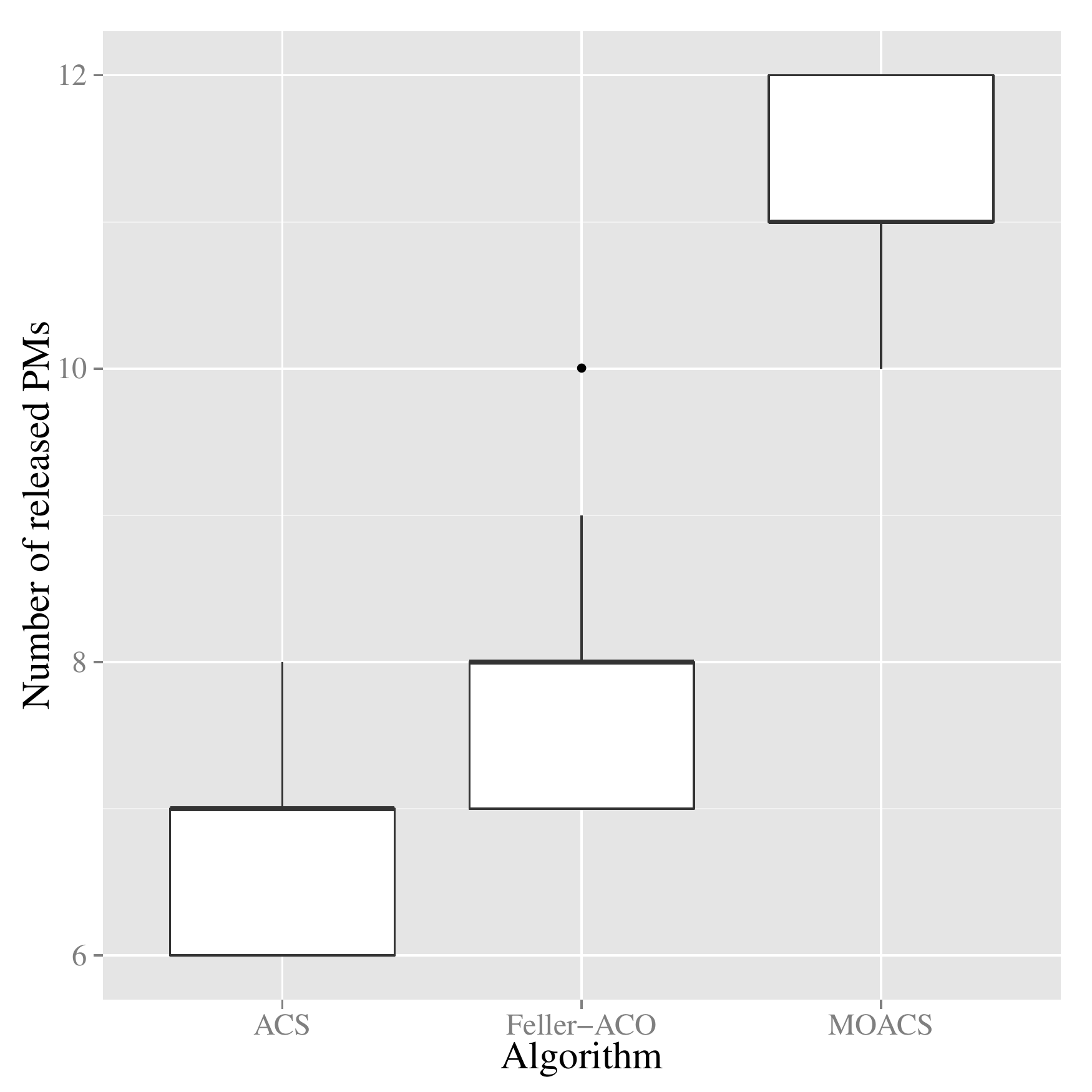}
        \caption{S2: High CPU, large memory}
    \end{subfigure}
  \caption{Number of released PMs}
    \label{fig:boxplots1}
\end{figure}

Figure~\ref{fig:boxplots1} presents the number of released PMs by the ACS, MOACS, and Feller-ACO solvers for the different scenarios as box plots. Moreover, Table~\ref{tab:summary-releases} provides a summary of the results in the numerical form. The table also provides the packing efficiency achieved by each solver. This variable is derived easily from the number of released PMs.

The results show that the MOACS solver was able to release 25\% to 37\% more PMs than the Feller-ACO solver, depending on the scenario. For example, in Scenario 1, MOACS released 15 PMs (median of 10 test runs) while Feller-ACO released only 11 PMs (median of 10 test runs). Since the packing efficiency is derived from the number of released PMs, it follows a similar trend. The difference in the number of released PMs between the MOACS and Feller-ACO solvers is statistically significant (Wilcoxon Signed-Rank Test, p-value = 0.005).

\subsubsection{Number of VM Migrations}
\begin{figure}[t!]
    \centering
    \begin{subfigure}[t]{0.45\textwidth}
        \centering
        \includegraphics[trim=0 0 0 0, clip, width=\textwidth]{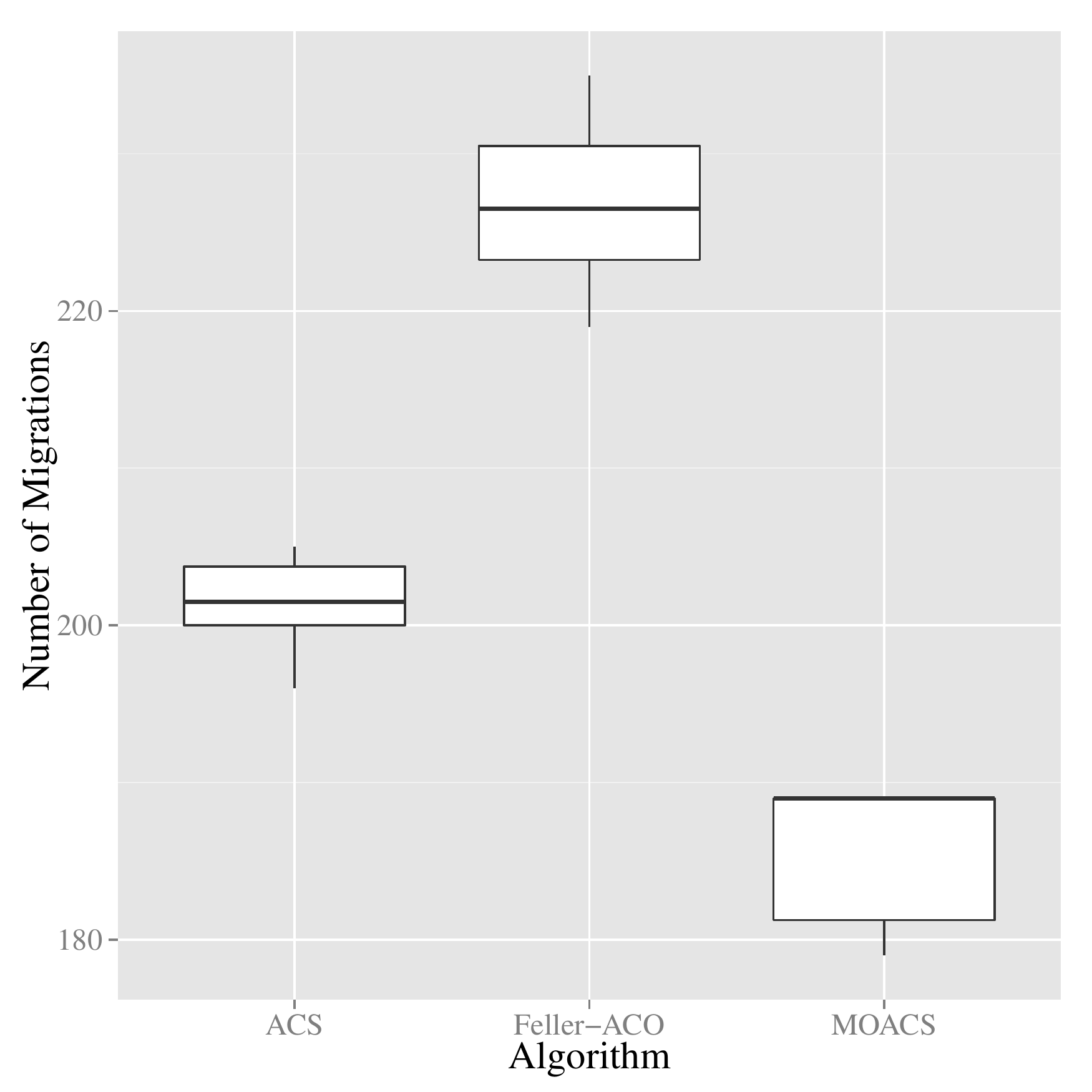}
        \caption{S1: Low CPU, small memory}
    \end{subfigure}
    \begin{subfigure}[t]{0.45\textwidth}
        \centering
        \includegraphics[trim=0 0 0 0, clip, width=\textwidth]{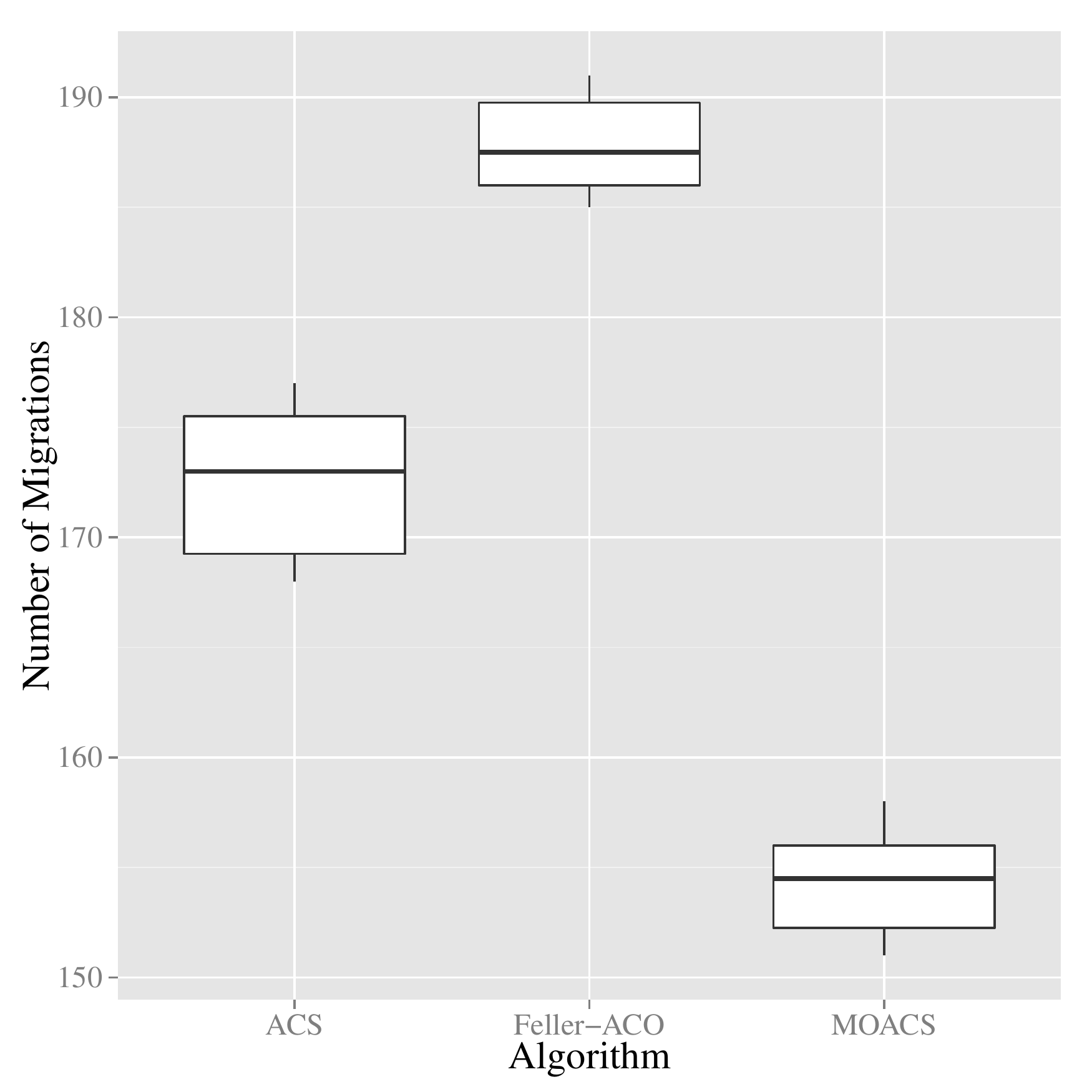}
        \caption{S3: High CPU, small memory}
    \end{subfigure}

    \begin{subfigure}[t]{0.45\textwidth}
        \centering
        \includegraphics[trim=0 0 0 0, clip, width=\textwidth]{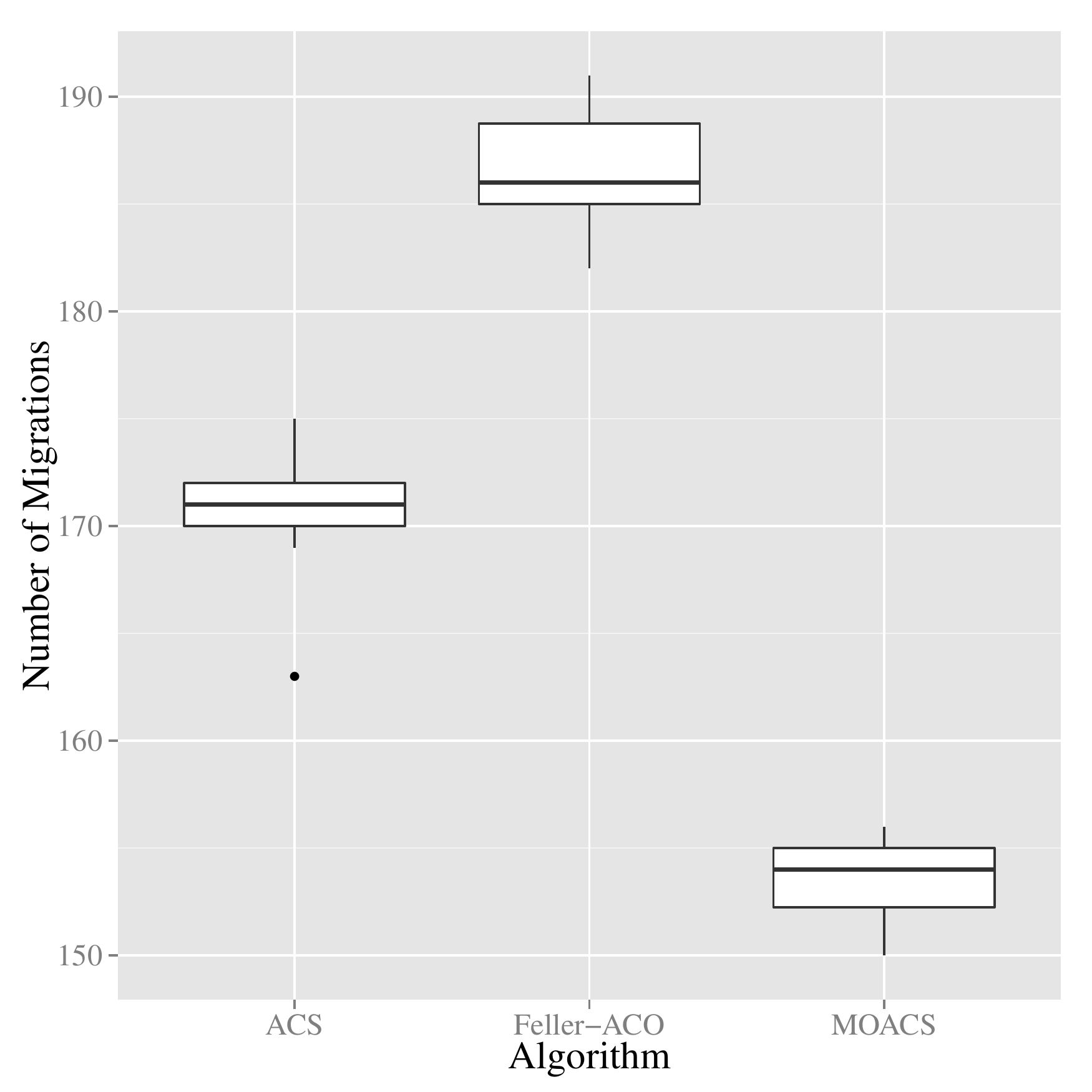}
        \caption{S4: Low CPU, large memory}
    \end{subfigure}
    \begin{subfigure}[t]{0.45\textwidth}
        \centering
        \includegraphics[trim=0 0 0 0, clip, width=\textwidth]{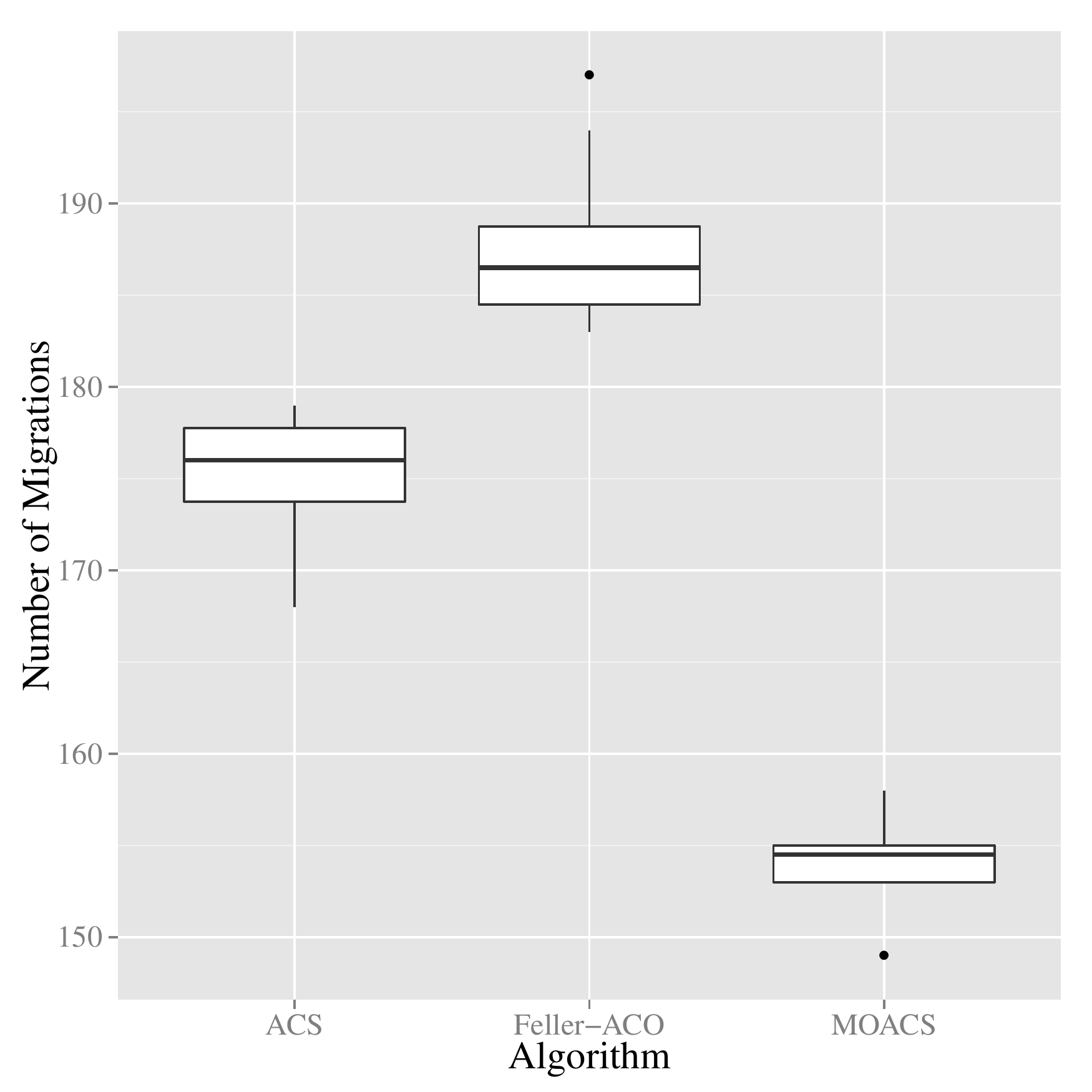}
        \caption{S2: High CPU, large memory}
    \end{subfigure}
  \caption{Number of VM migrations}
    \label{fig:boxplots2}
\end{figure}

The third dependent variable of interest in our experiment was the number of VM migrations, which should be minimized. Figure~\ref{fig:boxplots2} presents the results for this variable for the three solvers in the graphical form while Table~\ref{tab:summary-migrations} provides a summary of the results in the numerical form.

Again, we can observe that MOACS outperforms Feller-ACO for this objective. The results show that the MOACS solver was required to perform only 82\% to 83\% of the number of migrations required by the Feller-ACO solver to achieve an even better packing efficiency. For example, in Scenario 1, MOACS required 189 migrations (median of 10 test runs) while Feller-ACO required 226 migrations (median of 10 test runs). The difference in the number of migrations per solver is statistically significant (Wilcoxon Signed-Rank Test, p-value = 0.006).

\begin{table}[t]

\centering
\caption{Number of released PMs}
\label{tab:summary-releases}

\resizebox{\textwidth}{!}{%
\begin{tabular}{llrrrrr}
\toprule
                                                  &                                                                  & \multicolumn{1}{c}{ACS} & \multicolumn{1}{c}{\textbf{MOACS}} & \multicolumn{1}{c}{Feller-ACO} & \multicolumn{1}{c}{Change} & \multicolumn{1}{c}{p-value} \\ \hline
\multicolumn{1}{l}{\multirow{4}{*}{Scenario 1}} & Median Released                                                  & 9                        & 15                         & 11                          & 136\%                       & 0.005                        \\ \cline{2-7}
\multicolumn{1}{l}{}                            & Sd Released                                                      & 0.63                     & 0.52                       & 0.82                        &                             &                              \\ \cline{2-7}
\multicolumn{1}{l}{}                            & \begin{tabular}[c]{@{}l@{}}Packing \\ Efficiency \%\end{tabular} & 9\%                      & 15\%                       & 11\%                        & \multicolumn{1}{l}{}       & \multicolumn{1}{l}{}        \\ \hline
\multicolumn{1}{l}{\multirow{4}{*}{Scenario 2}} & Median Released                                                  & 7                        & 11                         & 8                           & 137\%                       & 0.005                        \\ \cline{2-7}
\multicolumn{1}{l}{}                            & Sd Released                                                      & 0.67                     & 0.67                       & 0.99                        &                             &                              \\ \cline{2-7}
\multicolumn{1}{l}{}                            & \begin{tabular}[c]{@{}l@{}}Packing \\ Efficiency \%\end{tabular} & 3.5\%                    & 5.5\%                      & 4\%                         & \multicolumn{1}{l}{}       & \multicolumn{1}{l}{}        \\ \hline
\multicolumn{1}{l}{\multirow{4}{*}{Scenario 3}} & Median Released                                                  & 7                        & 10                         & 8                           & 125\%                       & 0.004                        \\ \cline{2-7}
\multicolumn{1}{l}{}                            & Sd Released                                                      & 0.67                     & 0.52                       & 0.82                        &                             &                              \\ \cline{2-7}
\multicolumn{1}{l}{}                            & \begin{tabular}[c]{@{}l@{}}Packing \\ Efficiency \%\end{tabular} & 3.5\%                    & 5\%                        & 4\%                         & \multicolumn{1}{l}{}       & \multicolumn{1}{l}{}        \\ \hline
\multicolumn{1}{l}{\multirow{4}{*}{Scenario 4}} & Median Released                                                  & 7                        & 11                         & 8                           & 137\%                       & 0.005                        \\ \cline{2-7}
\multicolumn{1}{l}{}                            & Sd Released                                                      & 0.88                     & 0.88                       & 0.74                        &                             &                              \\ \cline{2-7}
\multicolumn{1}{l}{}                            & \begin{tabular}[c]{@{}l@{}}Packing \\ Efficiency \%\end{tabular} & 3.5\%                    & 5.5\%                      & 4\%                         & \multicolumn{1}{l}{}       & \multicolumn{1}{l}{}        \\
\botrule
\end{tabular}
}

\vspace{0.4cm}

\centering
\caption{Number of VM migrations}
\label{tab:summary-migrations}
\resizebox{\textwidth}{!}{
\begin{tabular}{llrrrrr}
\toprule
                                                  &                   & \multicolumn{1}{c}{ACS} & \multicolumn{1}{c}{\textbf{MOACS}} & \multicolumn{1}{c}{Feller-ACO} & \multicolumn{1}{c}{Change} & \multicolumn{1}{c}{p-value} \\ \hline
\multicolumn{1}{l}{\multirow{2}{*}{Scenario 1}} & Median Migrations & 201.5                    & 189                        & 226                         & 83\%                        & 0.002                        \\ \cline{2-7}
\multicolumn{1}{l}{}                            & Sd Migrations     & 3.13                     & 4.77                       & 5.23                        &                             &                              \\ \hline
\multicolumn{1}{l}{\multirow{2}{*}{Scenario 2}} & Median Migrations & 176                      & 154.5                      & 186                         & 83\%                        & 0.006                        \\ \cline{2-7}
\multicolumn{1}{l}{}                            & Sd Migrations     & 3.37                     & 2.31                       & 4.52                        &                             &                              \\ \hline
\multicolumn{1}{l}{\multirow{2}{*}{Scenario 3}} & Median Migrations & 173                      & 154.5                      & 187.5                       & 82\%                        & 0.005                        \\ \cline{2-7}
\multicolumn{1}{l}{}                            & Sd Migrations     & 3.34                     & 2.58                       & 2.35                        &                             &                              \\ \hline
\multicolumn{1}{l}{\multirow{2}{*}{Scenario 4}} & Median Migrations & 171                      & 154                        & 186                         & 83\%                       & 0.006                        \\ \cline{2-7}
\multicolumn{1}{l}{}                            & Sd Migrations     & 3.14                     & 2.07                       & 0.74                        &                             &                              \\
\botrule
\end{tabular}
}
\end{table}

\subsubsection{Execution Time and Scalability}
\begin{figure}[t!]
    \centering
    \begin{subfigure}[t]{0.9\textwidth}
      \centering
      \includegraphics[trim=0 0 0 0, clip,
      width=\textwidth]{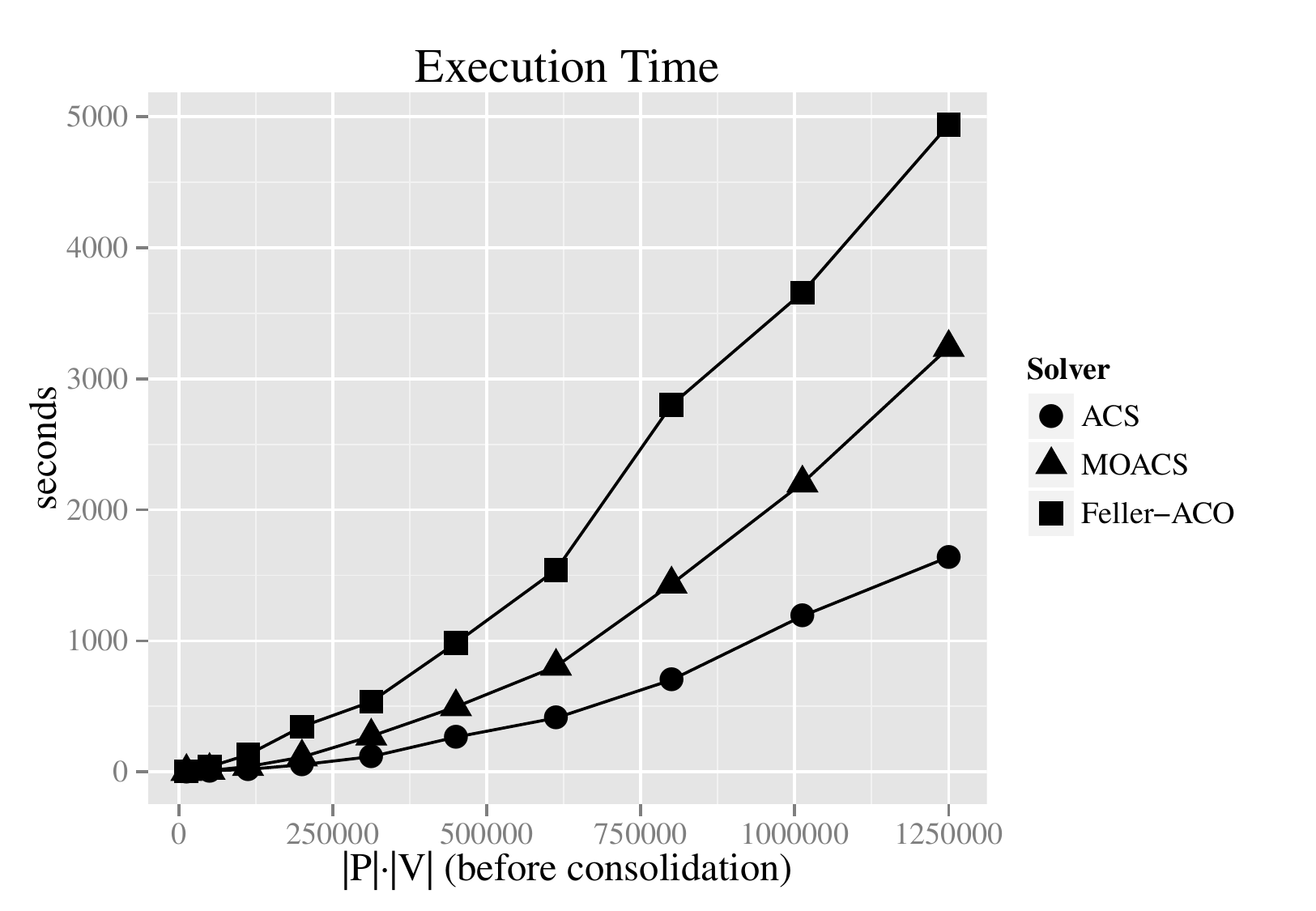}
    \end{subfigure}
 \caption{Scalability of the ACS, MOACS, and Feller-ACO solvers}
   \label{fig:time1}
 \end{figure}

The last comparison attribute is the execution time required by each solver to find a near-optimal, global best migration plan. Ideally, the solvers should use as less time as possible. Figure~\ref{fig:time1} presents the execution time for the ACS, MOACS, and Feller-ACO solvers when solving problems based on Scenario 2 with the number of PMs varying from 50 to 500 in increments of 50 and the number of VMs varying from 250 to 2500 in increments of 250. 

We can observe in the figure that MOACS performed better than Feller-ACO. Moreover, ACS performed better than MOACS but, to be fair, it used a single-objective algorithm while MOACS explored the search-space for two different objectives.

For reference, we also report the execution times to solve Scenario 1 with $|P|$=100 and $|V|$=1000 in Table~\ref{tab:time2}. For this scenario, ACS required a bit more than one minute. In contrast, MOACS required almost 2 minutes while Feller-ACO required almost 6 minutes. We report the median value for 10 test runs. However, we have observed that the standard deviation for the time variable was rather small for all solvers. The difference in the execution time between the MOACS and Feller-ACO solver is statistically significant (Wilcoxon Signed-Rank Test, p-value = 0.002).

\begin{table}[b!]
\caption{Execution time (in minutes) for the three solvers for Scenario 1}
   \label{tab:time2}
    \resizebox{\textwidth}{!}{

     \begin{tabular}{llrrrrr}
       \toprule
       &             & \multicolumn{1}{c}{ACS} & \multicolumn{1}{c}{\textbf{MOACS}} & \multicolumn{1}{c}{Feller-ACO} & \multicolumn{1}{c}{Speedup} & \multicolumn{1}{c}{p-value} \\ \hline
       \multicolumn{1}{l}{\multirow{2}{*}{$|P|\cdot|V|$=100000}} & Median Time & 1.03                     & 1.99                       & 5.92                        & 2.97X                        & 0.002                        \\ \cline{2-7}
       \multicolumn{1}{l}{}                            & Sd Time     & 0.03                     & 0.08                       & 0.17                        &                             &                              \\
       \botrule
     \end{tabular}
   }
\end{table}

\subsection{Analysis}

We can observe that the proposed MOACS algorithm and its corresponding solver outperformed Feller-ACO in all the measured variables: number of released PMs and packing efficiency (change 125\%), number of VM migrations (change 82\%), and execution time (speedup 2.97X). The solvers were exercised in four different scenarios involving different VM requirements. Each scenario was evaluated in 10 independent test runs. The differences were statistically significant for all variables (Wilcoxon Signed-Rank Test, p-values less than or equal to 0.006).

The differences in the performance can be explained by the design of each algorithm. Feller-ACO uses a single-objective, single-colony MMAS algorithm with an AOF that combines two different objectives concerning the number of released PMs and the number of VM migrations, whereas MOACS uses a multi-objective algorithm with two independent ant colonies for optimizing the two objectives. The AOF approach in Feller-ACO uses several parameters to determine the relative importance of the two objectives in the overall optimization process. We consider that this approach has two drawbacks: (1) it is difficult to find appropriate values for the different parameters in an AOF and (2) an AOF may not combine the different objectives in an appropriate manner. For instance, as described in Section~\ref{sec:moacs_wac}, maximizing the number of released PMs takes precedence in MOACS over minimizing the number of VM migrations. However, the AOF in Feller-ACO does not support precedence of one objective over another. Finally, MOACS uses additional constraints over its search-space, which significantly reduce the algorithm execution time without compromising on the quality of the solutions. The experimental evaluation clearly showed that these design decisions have an actual impact on the performance of the solvers.

It was also interesting to compare MOACS to ACS. MOACS clearly released more PMs and required less VM migrations than ACS. However, it was slower than ACS. When comparing ACS and Feller-ACO, we observed that ACS was faster and required less VM migrations than Feller-ACO, although it did not achieve the same packing efficiency. Still, ACS was the fastest of the three solvers and it can be a good alternative to consider when execution time is critical.

\section{Conclusion}
\label{conclusion}

We presented a novel multi-objective ant colony system algorithm for virtual machine (VM) consolidation in cloud data centers. The proposed algorithm builds VM migration plans, which are then used to reduce the number of required physical machines (PMs) by migrating and consolidating VMs on under-utilized PMs. It optimizes two objectives that are ordered by their importance. The first and foremost objective in the proposed algorithm is to maximize the number of released PMs. Moreover, since VM migration is a resource-intensive operation, it also tries to minimize the number of VM migrations.

The proposed algorithm was evaluated in a series of experiments. The experimental evaluation compared the proposed algorithm with two previously published ant colony optimization based VM consolidation algorithms, which were chosen for comparison due to their excellent performance with respect to different attributes. We considered four different scenarios of interest to test the three algorithms under different VM configurations. The experimental results showed that the proposed algorithm provided an efficient solution for VM consolidation in cloud data centers. Moreover, it outperformed the two existing ant colony optimization based VM consolidation algorithms in terms of number of released PMs, packing efficiency, and number of VM migrations.

\section*{Acknowledgements}
\noindent This work was supported by the Need for Speed (N4S) Research Program (\url{http://www.n4s.fi}).

\bibliographystyle{gPAA}
\bibliography{bibliography}

\end{document}